\documentclass[3p,preprint,times]{elsarticle}
\usepackage[T1]{fontenc}
\usepackage{amsmath,amsfonts}
\usepackage{mathtools}
\usepackage{IEEEtrantools}
\usepackage{graphicx}
\usepackage{tikz,pgfplots}
\usepackage{pgfplotstable}
\usepackage{tikzscale}
\usepackage{booktabs}
\usepackage{threeparttable}
\usepackage{multirow}
\usepackage{etoolbox}
\usepackage{acronym}
\usepackage[retain-unity-mantissa=false,exponent-product=\cdot,binary-units]{siunitx}
\usepackage{microtype}

\usepackage{mathrsfs}
\usepackage{bm}

\usetikzlibrary{%
  babel,%
  backgrounds,%
  calc,%
  decorations.markings,%
  external,%
  fit,%
  matrix,%
  patterns,%
  plotmarks,%
  positioning,%
  shapes.geometric%
}

\newcommand*{\plotratio}{1.618}

\makeatletter
  \RenewDocumentCommand{\tikzscale@prepareTikzpicture}{}{%
    \tikzset{
	  every picture/.style={%
	    scale=\tikzscale@scale,
		thin,
		every picture/.style={thin}
      }%
	}%
}%
\makeatother

\newlength{\mylinewidth}
\tikzset%
{%
  ultra thin/.style= {line width=0.25\mylinewidth},
  very thin/.style=  {line width=0.5\mylinewidth},
  thin/.style=       {line width=\mylinewidth},
  semithick/.style=  {line width=1.5\mylinewidth},
  thick/.style=      {line width=2\mylinewidth},
  very thick/.style= {line width=3\mylinewidth},
  ultra thick/.style={line width=4\mylinewidth}
}

\newcommand*{\plotfontsize}{\footnotesize}

\newlength{\marksize}
\definecolor{mycolor1}{rgb}{0.00000,0.44700,0.74100}
\definecolor{mycolor2}{rgb}{0.85000,0.32500,0.09800}
\definecolor{mycolor3}{rgb}{0.92900,0.69400,0.12500}
\definecolor{mycolor4}{rgb}{0.49400,0.18400,0.55600}
\definecolor{mycolor5}{rgb}{0.46600,0.67400,0.18800}
\definecolor{mycolor6}{rgb}{0.30100,0.74500,0.93300}
\definecolor{mycolor7}{rgb}{0.63500,0.07800,0.18400}

\pgfplotsset%
{%
  compat=1.14,
  plot coordinates/math parser=false,
  grid style={thin},
  minor grid style={black!10},
  every axis/.append style=
  {
    thin,
    mark options={style=solid, thin, mark size=\marksize},
    tick style={thin},
    label style={thin, font=\plotfontsize},
    tick label style={thin, font=\plotfontsize},
    legend style={thin, font=\plotfontsize, legend cell align=left},
    ylabel shift = -3pt,
    xlabel shift = -3pt
  },
  every picture/.style={thin},
  colormap={wygb}{rgb255=(255,255,255) rgb255=(255,255,204) rgb255=(161,218,180) rgb255=(65,182,196) rgb255=(34,94,168)},
  colormap={wgrayr}{rgb255=(255,255,255) color=(black!10) color=(mycolor2!70)},
  winDist/.style={surf, shader=flat corner, draw=black},
  winEs/.style={surf, shader=flat corner, draw=black, opacity=0.2, fill opacity=1.0}
}

\tikzset%
{%
  every label/.append style={thin, font=\plotfontsize},
  every node/.append style={thin, font=\plotfontsize},
  every edge/.append style={thin, font=\plotfontsize},
  dsp/.append style={thin, font=\plotfontsize},
  dsp/label/.append style={thin, font=\plotfontsize},
  dspfilter/.append style={thin, font=\plotfontsize},
  dspsquare/.append style={thin, font=\plotfontsize},
  dspconn/.append style={thin, font=\plotfontsize},
  dspline/.append style={thin, font=\plotfontsize}
}

\newlength{\tikzthin}
\newcommand*{\widthTwoThirds}{0.66667\linewidth}

\tikzset%
{%
  legendTransparent/.style={fill opacity=0.7, draw opacity=1, text opacity=1},
  blockZero/.style={draw=black!50},
  blockFull/.style={blockZero, fill=mycolor1!40},
  qcElement/.style={draw=black!50, fill=mycolor1!60},
  sbB/.style = {regular polygon,regular polygon sides=4,inner sep=0,outer sep=0,text width=width("0000"),minimum width=width("0000"),align=center,transform shape,font=\large},
  sbB1/.style = {sbB, draw, fill=mycolor1!5},
  sbB2/.style = {sbB, draw, fill=mycolor1!20},
  sbB3/.style = {sbB, draw, fill=mycolor1!35},
  sbC/.style = {regular polygon,regular polygon sides=4,inner sep=0,outer sep=0,text width=width("0000"),minimum width=width("0000"),align=center,transform shape,scale=0.4},
  sbC1/.style = {sbC, draw, pattern color=mycolor1, pattern=north east lines},
  sbC2/.style = {sbC, draw, pattern color=mycolor2, pattern=north west lines},
  sbC3/.style = {sbC, draw, pattern color=mycolor3, pattern=north east lines},
  sbC4/.style = {sbC, draw, pattern color=mycolor4, pattern=north west lines},
  sbC5/.style = {sbC, draw, pattern color=mycolor5, pattern=north east lines},
  sbC6/.style = {sbC, draw, pattern color=mycolor6, pattern=north west lines},
  harqNode0/.style={draw, fill=black!5},
  harqNode1/.style={draw, fill=mycolor2!50},
  harqCluster0/.style = {harqNode0},
  harqCluster1/.style = {draw, pattern=north east lines, pattern color=mycolor2},
  harqCluster1Inf/.style = {draw, pattern=north east lines, pattern color=mycolor1},
  harqCluster1Par/.style = {draw, pattern=north west lines, pattern color=mycolor4},
  edgeOut/.style = {draw, pattern=crosshatch, pattern color=black!25},
  edgeCla/.style = {pattern=north east lines, pattern color=mycolor1},
  edgeExt/.style = {pattern=north west lines, pattern color=mycolor2},
  setFut/.style = {draw, fill=black!10},
  setPre/.style = {draw, fill=mycolor2!30},
  setTar/.style = {draw, fill=mycolor1!60},
  setMid/.style = {draw, fill=mycolor1!40},
  setRem/.style = {draw, fill=mycolor1!20},
  setCore/.style = {draw, fill=mycolor1!15},
  setExt/.style  = {draw, fill=mycolor2!15},
  setPunc/.style = {draw, dotted, pattern=north east lines, pattern color=mycolor1!50},
  parityG/.style = {color=mycolor1},
  parityB/.style = {color=mycolor2},
  bus/.style = {postaction={decorate}, decoration={markings,mark= at position #1 with {\node[font=\footnotesize] {$\diagup$};}}}
}

\input{settings_other}
\acrodef{ACK}{Acknowledgment}
\acrodef{AMR}{Adapative Multi-Rate}
\acrodef{ANEU}{Average Number of Edge Updates} 
\acrodef{ANI}{Average Number of Iterations}
\acrodef{ANMU}{Average Number of Message Updates}
\acrodef{APP}{A-Posteriori Probability}
\acrodef{ARQ}{Automatic Repeat reQuest}
\acrodef{ASIC}{Application-Specific Integrated Circuit}
\acrodef{AWGN}{Additive White Gaussian Noise}
\acrodef{BEC}{Binary Erasure Channel}
\acrodef{BER}{Bit Error Ratio}
\acrodef{BD}{Boomerang Decoding}
\acrodef{BG}{Base Graph}
\acrodef{BLER}{Block Error Ratio}
\acrodef{BPR}{Bad-Parity Ratio}
\acrodef{BPSK}{Binary Phase-Shift Keying}
\acrodef{CBER}{Code-Block-Error Ratio} 
\acrodef{CN}{Check Node}
\acrodef{COPC}{Core-Only Parity Check}
\acrodef{CRC}{Cyclic Redundancy Check}
\acrodef{CSI}{Channel State Information}
\acrodef{DC}{Diversity Combining}
\acrodef{DS}{Dynamic Scheduling}
\acrodef{EGU}{Early Give-Up}
\acrodef{ES}{Early Success}
\acrodef{ET}{Early Termination}
\acrodef{FBD}{Full Block Decoder}
\acrodef{FC}{Forced Convergence}
\acrodef{FDC}{Full Diversity Combining}
\acrodef{FEC}{Forward Error Correction}
\acrodef{FPR}{False-Positive Rate}
\acrodef{HARQ}{Hybrid Automated Repeat reQuest}
\acrodef{IDS}{Informed Dynamic Scheduling}
\acrodef{IR}{Incremental-Redundancy}
\acrodef{LDPC}{Low-Density Parity-Check}
\acrodef{LLR}{Log-Likelihood Ratio}
\acrodef{MAP}{Maximum A-Posteriori}
\acrodef{MI}{Mutual Information}
\acrodef{MIMO}{Multiple-Input Multiple-Output}
\acrodef{MRC}{Maximum-Ratio Combining}
\acrodef{MSA}{Min-Sum Algorithm}
\acrodef{NAK}{Negative Acknowledgment}
\acrodef{NR}{New Radio}
\acrodef{OFDM}{Orthogonal Frequency-Division Multiplexing}
\acrodef{PBPR}{Parity-Based Partial-Retransmission}
\acrodef{PC}{Parity-Check}
\acrodef{PCM}{Parity-Check Matrix}
\acrodef{pdf}{probability density function}
\acrodef{PoD}{Probability of Dissatisfaction}
\acrodef{PSD}{Power Spectral Density}
\acrodef{QAM}{Quadrature Amplitude Modulation}
\acrodef{QC}{Quasi-Cyclic}
\acrodef{QPSK}{Quadrature Phase-Shift Keying}
\acrodef{RAC}{Retransmission-Aware Cost}
\acrodef{RB}{Reliability-Based}
\acrodef{RBP}{Residual Belief Propagation}
\acrodef{ROC}{Receiver Operating Characteristic}
\acrodef{SB}{Sub-Block}
\acrodef{SC}{Spatially-Coupled}
\acrodef{SCRB}{Standard Clustered Reliability-Based}
\acrodef{SFBC}{Space\textendash Frequency Block Coding}
\acrodef{SNR}{Signal-to-Noise Ratio}
\acrodef{SPA}{Sum-Product Algorithm}
\acrodef{SPCE}{Single-Parity-Check Extension}
\acrodef{TPR}{True-Positive Rate}
\acrodef{TtB}{Top-to-Bottom}
\acrodef{VN}{Variable Node}
\acrodef{WD}{Windowed Decoder}

\acrodefplural{APP}{A-Posteriori Probabilities}
\acrodefplural{PoD}{Probabilities of Dissatisfaction}

\acrodefindefinite{FC}{an}{a}
\acrodefindefinite{FBD}{an}{a}
\acrodefindefinite{FDC}{an}{a}
\acrodefindefinite{FPR}{an}{a}
\acrodefindefinite{HARQ}{an}{a}
\acrodefindefinite{LDPC}{an}{a}
\acrodefindefinite{LLR}{an}{a}
\acrodefindefinite{MAP}{an}{a}
\acrodefindefinite{MI}{an}{a}
\acrodefindefinite{MSA}{an}{a}
\acrodefindefinite{NR}{an}{a}
\acrodefindefinite{SB}{an}{a}
\acrodefindefinite{RAC}{an}{a}
\acrodefindefinite{RB}{an}{a}
\acrodefindefinite{RBP}{an}{a}
\acrodefindefinite{ROC}{an}{a}
\acrodefindefinite{SB}{an}{a}
\acrodefindefinite{SC}{an}{a}
\acrodefindefinite{SNR}{an}{a}
\acrodefindefinite{SPA}{an}{a}
\acrodefindefinite{SPCE}{an}{a}

\makeatletter
\newcommand{\acresetsingle}[1]{%
  \AC@reset{#1}%
}
\makeatother

\makeatletter
\@ifpackageloaded{unicode-math}{%
  \renewcommand*{\vec}[1]{\symbf{#1}}
}{%
  \renewcommand*{\vec}[1]{\bm{#1}}
}
\makeatother
\newcommand*{\transpose}{^{\mkern-1.5mu\mathsf{T}}}

\DeclareMathOperator{\sign}{sign}

\DeclareMathOperator{\lifting}{\mathscr{L}}

\DeclarePairedDelimiter{\curly}{\lbrace}{\rbrace}
\DeclarePairedDelimiter{\abs}{\lvert}{\rvert}

\DeclarePairedDelimiter{\floor}{\lfloor}{\rfloor}

\newcommand*{\subMax}{\text{max}}

\newcommand*{\subSymbol}{\text{s}}

\newcommand*{\subApp}{\text{a}}
\newcommand*{\subCB}{\text{c}}
\newcommand*{\subCn}{\text{c}}
\newcommand*{\subVn}{\text{v}}
\newcommand*{\subAll}{\text{a}}

\newcommand*{\subExt}{\text{e}}

\newcommand*{\subRem}{\text{r}}
\newcommand*{\subTime}{\text{s}}
\newcommand*{\subTerm}{\text{t}}

\newcommand*{\subTar}{\text{t}}
\newcommand*{\subCom}{\text{c}}
\newcommand*{\subMid}{\text{m}}
\newcommand*{\subMsg}{\text{m}}

\newcommand*{\subMessage}{\text{m}}

\newcommand*{\probSymbol}{p}
\newcommand*{\ProbSymbol}{P}
\newcommand*{\pdf}[2]{\probSymbol_{#1}(#2)}
\WithSuffix{\newcommand*}{\pdf}*[2]{\probSymbol_{#1}\left(#2\right)}

\newcommand*{\setInteger}{\mathbb{Z}}

\newcommand*{\rxSig}{y}
\newcommand*{\RxSig}{Y}

\newcommand*{\llr}{L}

\newcommand*{\llrChan}[1]{\llr_{\subApp,#1}}

\newcommand*{\energy}{E}
\newcommand*{\symbolEnergy}{\energy_\subSymbol}

\newcommand*{\psdWhite}{N_0}

\newcommand*{\esno}{\zeta}

\newcommand*{\rate}{R}

\newcommand*{\infoBit}{b}

\newcommand*{\parityBit}{p}
\newcommand*{\codeBit}{c}
\newcommand*{\CodeBit}{C}

\newcommand*{\numVn}{n}
\newcommand*{\numCn}{m}
\newcommand*{\numInfo}{k}

\newcommand*{\numVnSb}{N}
\newcommand*{\numCnSb}{M}
\newcommand*{\gf}{\mathbb{F}_2}

\newcommand*{\pcm}{\vec{H}}

\newcommand*{\liftFac}{\Theta}

\newcommand*{\expMatEl}{E}
\newcommand*{\expMat}{\vec{\expMatEl}}

\newcommand*{\idxVn}{i}
\newcommand*{\idxCn}{j}
\newcommand*{\msgCV}[2]{{\llr_\subCn(#1\to#2)}}
\newcommand*{\msgVC}[2]{{\llr_\subVn(#1\to#2)}}

\newcommand*{\setEdges}{\mathcal{E}}
\newcommand*{\setEdgesCn}[1]{{\setEdges_\subCn(#1)}}
\newcommand*{\setEdgesVn}[1]{{\setEdges_\subVn(#1)}}

\newcommand*{\degNode}{d}
\newcommand*{\degCn}{{\degNode_\subCn}}
\newcommand*{\degVn}{{\degNode_\subVn}}

\newcommand*{\bler}{\ProbSymbol_\subCB}

\newcommand*{\numIter}{\Lambda}

\newcommand*{\numMU}{I}
\newcommand*{\aneu}{\bar{\numMU}}
\newcommand*{\aneuRel}{\bar{\iota}}

\newcommand*{\layer}{\mathcal{L}}
\newcommand*{\idxLayer}{l}

\newcommand*{\numIterMax}{{\numIter_\subMax}}

\newcommand*{\numMUMax}{{\numMU_\subMax}}

\newcommand*{\mem}{\Psi}
\newcommand*{\idxMem}{\psi}
\newcommand*{\idxTime}{t}
\newcommand*{\period}{T}
\newcommand*{\constraintLength}{\nu}
\newcommand*{\numVnSbTime}{\numVnSb_\subTime}
\newcommand*{\numCnSbTime}{\numCnSb_\subTime}
\newcommand*{\numCouple}{J}
\newcommand*{\numCoupleUser}{\numCouple_\subMessage}
\newcommand*{\numCoupleTerm}{\numCouple_\subTerm}

\newcommand*{\winSize}{W}
\newcommand*{\idxWin}{\omega}
\newcommand*{\numWin}{\Omega}
\newcommand*{\win}[2]{\mathcal{W}_#1^{(#2)}}
\newcommand*{\setVnWin}{\mathcal{V}}
\newcommand*{\setCnWin}{\mathcal{C}}
\newcommand*{\setVnWinExt}{\setVnWin_\subExt}

\newcommand*{\setCnWinTar}{\setCnWin_\subTar}
\newcommand*{\setCnWinMid}{\setCnWin_\subMid}
\newcommand*{\setCnWinRem}{\setCnWin_\subRem}
\newcommand*{\setCnWinCom}{\setCnWin_\subCom}
\newcommand*{\setCnWinAll}{\setCnWin_\subAll}

\newcommand*{\numMsgWin}{\numVnSb_\subMsg}

\newcommand*{\cfg}{\mathcal{U}}
\newcommand*{\cfgCN}{\cfg_\text{CN}}
\newcommand*{\cfgVN}{\cfg_\text{VN}}

\renewcommand*{\idxMem}{\mu}
\renewcommand*{\mem}{{m_\subTime}}
\renewcommand*{\constraintLength}{\nu_\subTime}
\renewcommand*{\numCouple}{L}
\renewcommand*{\liftFac}{Q}
\renewcommand*{\numWin}{\varOmega}
\renewcommand*{\idxWin}{w}
\renewcommand*{\esno}{\symbolEnergy/\psdWhite}
\renewcommand*{\subMessage}{\text{u}}

\AtBeginEnvironment{tabular}{\footnotesize}
\AtBeginEnvironment{tabularx}{\footnotesize}
\AtBeginEnvironment{tablenotes}{\footnotesize}

\begin{document}

\begin{frontmatter}

\title{Comparison of Windowed-Decoder Configurations for Spatially Coupled LDPC Codes Under Equal-Complexity Constraints}

\author[1]{Janik~Frenzel, M.Sc.}
\author[2]{Dr.-Ing.\ Stefan~M\"uller-Weinfurtner}
\author[3]{Dr.-Ing.\ Johannes~Huber}
\author[3]{Dr.-Ing.\ Ralf~M\"uller}

\address[1]{Intel, S\"udwestpark 2\textendash 4, 90449, N\"urnberg, Germany}
\address[2]{Cisco, Nordostpark 12, 90411 N\"urnberg, Germany}
\address[3]{Friedrich-Alexander-Universit\"at Erlangen-N\"urnberg, Digital Communications, Cauerstr.\ 7, 91058 Erlangen}

\begin{abstract}
  Spatially Coupled Low-Density Parity-Check (SC-LDPC) codes offer excellent decoding performance and can be elegantly decoded with a Windowed Decoder (WD). We determine an efficient WD configuration with low control overhead. For fair comparisons, we normalize all configurations to the same maximal computational complexity, which is an important measure of the decoding effort in packet-based data communication systems. We determine an optimized configuration from a joint evaluation of the window size, the window update strategy, and parity check\textendash based Early Termination (ET). Firstly, we use a variable node\textendash centered update strategy, which omits updates of messages in some parts of the decoding window. With the complexity normalization, the window size can be increased compared to a check node\textendash centered update strategy, which uniformly updates all messages in the decoding window. Secondly, we only require the satisfaction of the top-most parity-check equations in each window to move to the next position more quickly. Using a surprisingly large window size, the resulting WD halves the average decoding complexity of the block decoder while maintaining a rather small gap in the decoding performance.
\end{abstract}


\end{frontmatter}

\section{Introduction}
\label{sec:intro}
\ac{SC} \ac{LDPC} codes with \ac{QC} properties~\cite{JimenezFeltstrom1999,Tanner2004} present an appealing alternative to \ac{LDPC} block codes because of their compact representation and excellent performance: A single \ac{SC}-\ac{LDPC} code ensemble can universally achieve the capacity of a wide range of channels~\cite{Kudekar2013}.

\acused{WD}
There are decoder designs in literature that are more adapted to the convolutional structure of \ac{SC}-\ac{LDPC} codes than \iac{FBD}, which treats the code as a block code. On the one hand, a \emph{pipeline decoder} can utilize multiple processors to perform successive iterations on different subsets of the code's Tanner graph~\cite{JimenezFeltstrom1999}. On the other hand, a \acfi{WD}\textemdash also called \emph{sliding-window decoder}\textemdash uses a single processor to perform multiple iterations on the same subset of the graph~\cite{Pusane2008,Iyengar2012}. When reliability conditions are fulfilled or a maximal number of iterations is exhausted, the window slides to the next position. The decoding process resumes with some new information as windows at two successive positions overlap to a large extent.

We assume strict resource constraints to target low-power \ac{ASIC} implementations and thus prefer the single-instance \ac{WD} over the pipeline decoder. Additionally, the \ac{WD} has two inherent advantages over the \ac{FBD}: Firstly, the required amount of memory\textemdash which is typically rather expensive in \ac{ASIC} designs\textemdash is greatly reduced. Secondly, the special structure of \ac{SC}-\ac{LDPC} codes makes it possible to detect decoding errors at an early stage. As the decoding window usually moves unidirectionally, no further message updates are scheduled in parts of the Tanner graph that have already been processed by the decoder. Residual decoding errors in the related code symbols cannot be corrected afterwards. Alternatively, there exist proposals on special \ac{WD} designs where the decoding window moves bidirectionally to prevent the decoder from getting stuck~\cite{Abu-Surra2015,Klaiber2018}. However, these designs require additional control loops and are not further considered in this paper.

\acused{ET}
With the restrictions to unidirectional decoding and low-overhead procedures to simplify hardware implementations, there remain three major aspects that determine the decoding performance and the computational complexity of a \ac{WD}: Firstly, the \emph{size of the decoding window} to control the amount of information processed at once. Secondly, the \emph{update strategy} within the window to control which parts of the decoding window are updated. Thirdly, \acfi{ET} to stop the decoding before a maximal number of iterations is exhausted, thereby reducing the computational complexity.

One update strategy is to specify the window by a set of \acp{VN} that are updated~\cite{Pusane2008,Papaleo2010,UlHassan2017}. However, some \acp{VN} within a window then share \acp{CN} with \acp{VN} that have already moved out of that window. Messages sent along the corresponding edges are not updated any longer, i.e., they are read-only. There exist proposals on overcoming this potential drawback, e.g., by applying some form of amplification to these read-only messages~\cite{Ali2017}. A different update strategy inherently avoids this issue by defining the window by a set of \acp{CN} that are updated~\cite{Iyengar2012,Beemer2016}. With this strategy, messages sent along all edges connected to the involved \acp{CN} receive updates, including the messages omitted in the previously described update strategy. This \ac{CN}-centered procedure nicely matches \ac{CN}-wise serial update schedules that generate all outgoing messages from a \ac{CN} at the same time and do so one \ac{CN} after the other~\cite{Yeo2001}. Such update schedules work well with the \ac{MSA} typically implemented in practice since the minimum must be found across all messages directed towards each \ac{CN}~\cite{Fossorier1999}. Still, we also evaluate the \ac{VN}-centered procedure with \ac{CN}-wise serial update schedules for reasons of fairness.

For \ac{ET}, there again exist at least two distinct approaches. Many criteria in literature are based on code symbol reliabilities~\cite{UlHassan2017,Mo2017}. Alternatively, the \ac{PC} equations inherent to \ac{SC}-\ac{LDPC} codes can be used for \ac{ET} just as for \ac{LDPC} block codes: The decoding window moves to the next position as soon as all \ac{PC} equations in the window are fulfilled~\cite{Pusane2008}. We favor \ac{PC}-based criteria because the most recent \ac{PC} results are implicitly contained in the signs of the messages exchanged during iterative decoding when using \acp{LLR} and \ac{CN}-wise update scheduling. Hence, there is no computational overhead as the soft-value processing required for the \acp{LLR}, e.g., combination of multiples values and comparisons against a threshold, can be omitted. Still, the question remains whether \emph{all} \ac{PC} equations in the window should be considered for \ac{ET} or just a subset of them. Although partial evaluations are proposed in~\cite{Abu-Surra2015,Kang2018}, to our best knowledge no evaluation of the computational complexity for soft-input soft-output decoding is available so far.

In sum, a \ac{WD} can be configured in different ways than \iac{FBD}. Still, to our best knowledge, no holistic evaluation of the window size, window update strategies, and \ac{PC}-based \ac{ET} is available in literature so far. We try to fill this gap by performing comparisons in terms of the decoding performance and, more importantly, the resulting computational complexity. The goal of this paper is to compare configurations of a \ac{WD} to find a low-complexity configuration with minimal control overhead. We find that the decoding performance of a \ac{WD} with the configuration determined in this paper is not far off from that of the \iac{FBD} while resulting in a significantly reduced computational complexity.

Lastly, we use terminated codes\textemdash which are used in practical systems\textemdash instead of continuous encoding and decoding. The termination also results in a superior decoding performance~\cite{Lentmaier2010}. We furthermore assume the use case of data packets, which usually requires all information symbols to be correctly estimated before further processing by higher layers of the communication protocol takes place. Thus, we ignore the inherent advantage in decoding latency a \ac{WD} has compared to \iac{FBD}. We rather consider the time or effort it takes to decode all information symbols to measure the decoding complexity. Consequently, we limit all decoding schemes evaluated in this paper to the (approximately) same maximal number of updates of messages along the edges in the Tanner graph of the code.

The remainder of this paper is structured as follows: Section~\ref{sec:bg} summarizes \ac{SC}-\ac{LDPC} codes and iterative message-passing algorithms; Section~\ref{sec:config} reviews \ac{WD} configurations. The setup for numerical results is listed in Section~\ref{sec:setup}. Numerical decoding results are presented in Section~\ref{sec:results}. The paper concludes with Section~\ref{sec:conclusion}.

\section{Background}
\label{sec:bg}

This section briefly recapitulates \ac{SC}-\ac{LDPC} codes with \ac{QC} properties and the principle of iterative decoding using message-passing algorithms. Even though only a particular set of codes is used in this paper, the concepts discussed in Section~\ref{sec:config} are not restricted to this special code design.

\subsection{Spatially-Coupled Quasi-Cyclic LDPC Codes}

We use asymptotically regular \ac{SC}-\ac{LDPC} codes with \ac{QC} properties~\cite{Mitchell2010}. \ac{QC} codes offer a variety of benefits, namely compact representation and great potential for parallelization~\cite{Kou2001}. Together with the sparseness of \ac{LDPC} codes enabling efficient encoding and iterative decoding, these beneficial properties are the reason \ac{QC}-\ac{LDPC} codes are used in several applications, e.g., the 5G \ac{NR} cellular-communications standard~\cite{TS38212}. Although \ac{QC}-\ac{LDPC} codes have originally been introduced by \emph{copy-and-permute} operations on graphs~\cite{Thorpe2003}, we prefer the matrix notation used in, e.g.,~\cite{SehoMyung2006}. Throughout this paper, bold symbols indicate vectors (noncapitalized letters) or matrices (capitalized letters).

\subsubsection{Quasi-Cyclic Low-Density Parity-Check Codes}

The binary \ac{PCM} $\pcm$ of a \ac{QC}-\ac{LDPC} code is constructed by \emph{lifting} an \emph{exponent matrix} $\expMat$ of size $\numCnSb\times\numVnSb$ where $\numCnSb$ and $\numVnSb$ denote the number of blocks of \acp{CN} and \acp{VN} in $\pcm$, respectively. With $\setInteger$ denoting the set of integers and $\gf=\{0,1\}$ denoting the binary field, we write
\begin{IEEEeqnarray}{rCl}
  \lifting: \expMat \in \setInteger^{\numCnSb\times\numVnSb} \mapsto \pcm \in \gf^{\numCnSb\liftFac \times \numVnSb\liftFac}.
\end{IEEEeqnarray}
Each element $\expMatEl_{\idxCn, \idxVn}$ of $\expMat$, $\idxCn=0\dots(\numCnSb-1)$, $\idxVn=0\dots(\numVnSb-1)$, denotes a \emph{cyclic shift}. The lifting operation $\lifting$ replaces the cyclic shifts $\expMatEl_{\idxCn, \idxVn}$ by square matrices of size $\liftFac\times\liftFac$ where $\liftFac$ is the \emph{lifting factor}. If $\expMatEl_{\idxCn, \idxVn}\ge0$, the substituted matrix is an identity matrix that has its rows circularly rotated to the right $\expMatEl_{\idxCn, \idxVn}$ times.\footnote{The term \emph{exponent matrix} comes from an alternative representation of the rotation by using an appropriate rotation matrix, which is multiplied with itself $\expMatEl_{\idxCn, \idxVn}$ times.} Otherwise, the substituted matrix is an all-zero matrix. The resulting \ac{PCM} $\pcm$ of the binary ($\numVn$, $\numCn-\numVn$) \ac{LDPC} code thus has $\numCn=\numCnSb\liftFac$ rows and $\numVn=\numVnSb\liftFac$ columns. Each $\liftFac\times\liftFac$ sub-matrix corresponds to $\liftFac$ consecutive \acp{CN} and \acp{VN}, forming the respective \ac{CN} and \ac{VN} \emph{blocks}.

\subsubsection{Terminated {QC}-{SC}-{LDPC} Codes}

Instead of transmitting codewords individually with respect to the exponent matrix $\expMat$, several codewords can be \emph{spatially coupled}. Thereby, a structure similar to convolutional codes is obtained~\cite{Papaleo2010}. The resulting code with memory $\mem$ is described by exponent matrices $\expMat_\idxMem(\idxTime)$ with $\idxMem=0\dots\mem$ that define \emph{sub-codes}. The argument $\idxTime$ denotes the time index for possibly time-varying sub-codes. The $\expMat_\idxMem(\idxTime)$ have size $\numCnSbTime\times\numVnSbTime$ each, i.e., $\numVnSbTime$ blocks of code bits are transmitted at each time index, forming $\numCnSbTime$ blocks of \ac{PC} equations. For a code with memory $\mem$, up to $\mem+1$ sub-codes $\expMat_\idxMem(\idxTime)$ influence each encoded code bit. The length of the entire code is defined by the \emph{coupling length} $\numCouple$. As argued above, we focus on terminated codes and thus use a finite coupling length. The entire exponent matrix $\expMat_{[0, \numCouple-1]}$ of a terminated \ac{SC}-\ac{LDPC} code with coupling length $\numCouple$ has $\numCnSb=\numCnSbTime(\numCouple+\mem)$ rows and $\numVnSb=\numVnSbTime\numCouple$ columns and is described by
\begin{IEEEeqnarray}{rCl}
  \expMat_{[0, \numCouple-1]} =
  \begin{bmatrix}
    \expMat_0(0)       &        &                                 \\
    \vdots             & \ddots &                                 \\
    \expMat_\mem(\mem) & \ddots & \expMat_0(\numCouple-1)         \\
                       & \ddots & \vdots                          \\
                       &        & \expMat_\mem(\numCouple+\mem-1) \\
  \end{bmatrix}.
  \label{eq:bg:pcm_term}
\end{IEEEeqnarray}
A code is said to be periodic with period $\period$ if $\expMat_\idxMem(\idxTime)=\expMat_\idxMem(\idxTime+\period) \;\forall \idxTime \;\forall \idxMem=0\dots\mem$. The code's \emph{constraint length} $\constraintLength = \numVnSbTime\liftFac(\mem+1)$ corresponds to the maximal width of the support of the rows of the entire \ac{PCM} $\pcm_{[0, \numCouple-1]}=\lifting\{\expMat_{[0, \numCouple-1]}\}$. With $(\cdot)\transpose$ denoting the transpose operation,\footnote{Many coding theorists prefer row vectors over column vectors for codewords. Recent {3GPP} releases however use column vectors for codewords, which is why we follow this way of notation.} a valid codeword $\vec{\codeBit}$ comprises the concatenation of sub-codewords $\vec{\codeBit}(\idxTime) \in \gf^{\numVnSbTime\liftFac\times1}$ such that
\begin{IEEEeqnarray}{rCl}
  \vec{\codeBit}\transpose = \left[ \, \vec{\codeBit}\transpose(0) ,\, \dots ,\, \vec{\codeBit}\transpose(\numCouple-1) \, \right]
  \label{eq:bg:codeword}
\end{IEEEeqnarray}
and fulfills (with the operators $\oplus$ and $\odot$ denoting addition and matrix\textendash vector multiplication in $\gf$, respectively)
\begin{IEEEeqnarray}{rCl}
  \bigoplus_{\idxMem=0}^{\mem} \pcm_\idxMem(\idxTime) \odot \vec{\codeBit}(\idxTime-\idxMem) = \vec{0} \;\forall \idxTime
  \label{eq:bg:conv}
\end{IEEEeqnarray}
where $\pcm_\idxMem(\idxTime)=\lifting\{\expMat_\idxMem(\idxTime)\}$. In~\eqref{eq:bg:conv}, handling of the termination is omitted for simplicity.

\subsubsection{Effects of the Termination}

The top-most and bottom-most $\mem\numCnSbTime$ \ac{CN} blocks in $\pcm_{[0, \numCouple-1]}$ have smaller degrees than the ones in the middle of the \ac{PCM}. This irregularity is the main reason for the excellent performance of \ac{SC}-\ac{LDPC} codes~\cite{Lentmaier2010}. Since \ac{SC}-\ac{LDPC} codes exhibit recursive encoder structures, some effort is required to determine the termination sequence (given by the last sub-codewords in~\eqref{eq:bg:codeword}) that fulfills the \ac{PC} equations located at the bottom of the \ac{PCM}~\cite{Chen2008}. By contrast, the fulfillment of the \ac{PC} equations at the top of the \ac{PCM} is trivially solved by initializing the encoder with an all-zero state.

Furthermore, termination reduces the code rate. Let $\numCoupleUser$ denote the number of coupling instants used to transmit user data (i.e., the information symbols) and $\numCoupleTerm$ denote the number of coupling instants used for the termination sequence. Then $\numCouple = \numCoupleUser + \numCoupleTerm$ and the length of the whole codeword is given by $\numVn = \numCouple\numVnSbTime\liftFac$; the number of information symbols is given by $\numInfo = \numCoupleUser(\numVnSbTime-\numCnSbTime)\liftFac$. Hence, the effective code rate is
\begin{IEEEeqnarray}{rCl}
  \rate_\numCouple = \frac{\numInfo}{\numVn} = \frac{\numCoupleUser\left( \numVnSbTime-\numCnSbTime \right)\liftFac}{\left( \numCoupleUser+\numCoupleTerm \right) \numVnSbTime\liftFac} = \frac{\numCoupleUser}{\numCoupleUser+\numCoupleTerm}\rate_\infty
  \label{eq:code:rate}
\end{IEEEeqnarray}
with the asymptotic code rate
\begin{IEEEeqnarray}{rCl}
  \rate_\infty = \lim_{\numCouple\to\infty} \rate_\numCouple = \frac{\numVnSbTime - \numCnSbTime}{\numVnSbTime}.
\end{IEEEeqnarray}

\subsubsection{Code Employed in This Paper}

We use the \ac{QC} code ensembles described in~\cite{Chandrasetty2014}. The codes are \emph{asymptotically regular}, i.e., all \acp{VN} have the same degree $\degVn$ and all \acp{CN} the same degree $\degCn$ if the termination is not considered~\cite{Mitchell2010}. Such an ensemble is in short denoted as a $(\degVn,\degCn)$ code ensemble. Theoretic analyses for these ensembles (albeit not requiring \ac{QC} sub-matrices) such as decoding thresholds are available in, e.g.,~\cite{Lentmaier2010}. We focus on codes with $\rate_\infty=1/2$, i.e., $\degCn=2\degVn$. We use the smallest possible parameters to achieve that rate: $\numVnSbTime=2$, $\numCnSbTime=1$. In addition, we require fully populated matrices $\expMat_\idxMem(\idxTime)$ for all $\idxMem=0\dots\mem$ such that $\mem=\degVn-1$ since $\numCnSbTime=1$. The termination sequence comprises $\numCoupleTerm=\mem$ sub-codewords $\vec{\codeBit}(\idxTime)$~\cite{Chen2008}. However, simulations are performed using only the all-zero codeword to avoid expensive matrix inversion. The codes are periodic with $\period=3$ such that reasonably large girths can be achieved according to the results in~\cite{Chandrasetty2014}. An ensemble is formed by independently and randomly choosing the cyclic shifts for each $\expMat_\idxMem(\idxTime)$ from a uniform distribution with support $[0, \liftFac-1]$ for each transmission. Finally, realizations where cycles of length 4 appear in the Tanner graph are excluded from the evaluation.

With respect to~\eqref{eq:bg:codeword}, the sub-codewords $\vec{\codeBit}(\idxTime) \in \gf^{\numVnSbTime\liftFac\times1}$ at each coupling instant $\idxTime=0\dots(\numCoupleUser-1)$ are formed in the following way (for $\numVnSbTime=2$ and $\numCnSbTime=1$, i.e., $\numVnSbTime-\numCnSbTime=1$):
\begin{IEEEeqnarray}{rCl}
  \vec{\codeBit}\transpose(\idxTime) = \left[ \, \vec{\infoBit}\transpose(\idxTime) ,\, \vec{\parityBit}\transpose(\idxTime) \, \right].
  \label{eq:bg:info_parity}
\end{IEEEeqnarray}
The $\vec{\infoBit}(\idxTime) \in \gf^{\liftFac\times1}$ contain information symbols whereas the $\vec{\parityBit}(\idxTime) \in \gf^{\liftFac\times1}$ contain parity symbols. In the termination sequence, the $\vec{\codeBit}(\idxTime)$ only consist of parity symbols. The concatenated vector
\begin{IEEEeqnarray}{rCl}
  \vec{\infoBit}\transpose = \left[ \, \vec{\infoBit}\transpose(0) ,\, \dots ,\, \vec{\infoBit}\transpose(\numCoupleUser-1) \, \right]
  \label{eq:bg:info_all}
\end{IEEEeqnarray}
then contains all $\numInfo$ information symbols.

\subsection{Message-Passing Decoding}

Although \ac{SC}-\ac{LDPC} codes can be decoded like regular convolutional codes, i.e., with a Viterbi decoder, their large constraint lengths render such an approach infeasible~\cite{Papaleo2010}. Instead, the sparseness of the $\pcm_\idxMem(\idxTime)$ enables efficient iterative decoding with belief propagation~\cite{MacKay1999}. The decoding output is given in the form of hard decisions that estimate the transmitted symbols from~\eqref{eq:bg:info_parity}:
\begin{IEEEeqnarray}{rCl}
  \vec{\hat{\codeBit}}\transpose(\idxTime) = \left[ \, \vec{\hat{\infoBit}}\transpose(\idxTime) ,\, \vec{\hat{\parityBit}}\transpose(\idxTime) \, \right].
  \label{eq:bg:info_parity_est}
\end{IEEEeqnarray}
For iterative decoding, the reliabilities of the $\numVn$ symbols of the complete received word $\vec{\rxSig}$ are usually represented by \acfp{LLR} with respect to the binary code symbols $\codeBit_\idxVn$, $\idxVn=0\dots(\numVn-1)$, to improve the numerical stability of the iterative decoding process. Let (without loss of generality) the code symbol $\codeBit_\idxVn$ be associated with a received symbol $\rxSig$. Then
\begin{IEEEeqnarray}{rCl}
  \llrChan{\idxVn} = \log \frac{\Pr\{ \CodeBit_\idxVn = 0 \mid \RxSig \}}{ \Pr\{ \CodeBit_\idxVn = 1 \mid \RxSig \} }
  \label{eq:bg:llr}
\end{IEEEeqnarray}
represents the \ac{LLR} given the received value where the random variables $\CodeBit_\idxVn$ and $\RxSig$ correspond to the realizations $\codeBit_\idxVn$ and $\rxSig$, respectively. The updated reliabilities (incorporating extrinsic information) $\llr_\idxVn$ for the $\numVn$ code symbols, i.e., $\idxVn=0\dots(\numVn-1)$, are calculated at the \acp{VN} with
\begin{IEEEeqnarray}{rCl}
  \llr_\idxVn = \llrChan{\idxVn} + \smashoperator[r]{\sum_{\idxCn \in\setEdgesCn{\idxVn}}} \msgCV{\idxCn}{\idxVn}
  \label{eq:bg:spa_rel}
\end{IEEEeqnarray}
where $\setEdgesCn{\idxVn}$ is the set of \acp{CN} connected to \ac{VN} $\idxVn$ and $\msgCV{\idxCn}{\idxVn}$ is the message from \ac{CN} $\idxCn$ to \ac{VN} $\idxVn$~\cite{Hagenauer1996}. The message sent from \ac{VN} $\idxVn$ to \ac{CN} $\idxCn$ in the next iteration is given by
\begin{IEEEeqnarray}{rCl}
  \msgVC{\idxVn}{\idxCn} = \llr_\idxVn - \msgCV{\idxCn}{\idxVn}.
  \label{eq:bg:spa_v2c}
\end{IEEEeqnarray}
In the \ac{SPA}, the \ac{CN} messages are computed from $\setEdgesVn{\idxCn}$, the set of \acp{VN} connected to \ac{CN} $\idxCn$, in an optimal way~\cite{Hagenauer1996}:
\begin{IEEEeqnarray}{rCl}
  \tanh\left( \frac{\msgCV{\idxCn}{\idxVn}}{2} \right) = \smashoperator[r]{\prod_{\idxCn \in\setEdgesVn{\idxCn} \setminus \idxVn}} \tanh\left( \frac{\msgVC{\idxVn}{\idxCn}}{2} \right).
  \label{eq:bg:spa_c2v}
\end{IEEEeqnarray}
Equation~\eqref{eq:bg:spa_c2v} is prohibitively complex for implementation in practice. Therefore, approximations like the \ac{MSA} are often applied~\cite{Fossorier1999}. In its simplest form, the \ac{MSA} replaces~\eqref{eq:bg:spa_c2v} by
\begin{IEEEeqnarray}{rCl}
  \msgCV{\idxCn}{\idxVn} &=& \smashoperator[r]{\prod_{\idxCn \in\setEdgesVn{\idxCn} \setminus \idxVn}} \sign\{\msgVC{\idxVn}{\idxCn}\} \cdot \min_{\idxCn \in\setEdgesVn{\idxCn} \setminus \idxVn} \bigl\{ \abs{\msgVC{\idxVn}{\idxCn}} \bigr\}
  \label{eq:bg:msa_c2v}
\end{IEEEeqnarray}
where the sign function is defined as
\begin{IEEEeqnarray}{rCl}
  \sign\{x\}=
  \begin{cases}
    0         &\text{if~} x=0,\\
    x/\abs{x} &\text{otherwise}.
  \end{cases}
\end{IEEEeqnarray}

\section{Windowed-Decoder Configurations Selected for Comparisons}
\label{sec:config}
\acresetsingle{ET}
\acresetsingle{FBD}

We briefly introduce our notation for the \ac{WD} and review the configurations relevant to the comparisons in this paper, i.e., window update strategies and \ac{PC}-based \ac{ET}.

\subsection{Window Update Strategies}

We need to strictly differentiate the \emph{window update strategy}, i.e., how the definition of the decoding window affects \ac{VN} and \ac{CN} updates, from \emph{update scheduling}, i.e., whether message updates are performed in a serial or parallel manner. In essence, two update strategies can be distinguished. For the first strategy\textemdash called the \emph{\ac{VN}-centered} strategy $\cfgVN$ from here on\textemdash a decoding window of size $\winSize$ is defined by the set of \acp{VN} for $\winSize$ consecutive sub-codewords $\vec{\codeBit}(\idxTime)$, cf.~\cite{Pusane2008,Papaleo2010,UlHassan2017}. When \ac{QC} codes as described above are used, the window thus consists of $\winSize\numVnSbTime$ consecutive \ac{VN} blocks, each of which consists of $\liftFac$ consecutive \acp{VN}. For the second strategy\textemdash called the \emph{\ac{CN}-centered} strategy $\cfgCN$ from here on\textemdash a decoding window of size $\winSize$ is defined by the set of $\winSize\numCnSbTime$ consecutive \ac{CN} blocks and all related \acp{VN}~\cite{Iyengar2012,Beemer2016}.

\begin{figure}[htb]
  \centering
  \tikzsetnextfilename{config_window}
  \tikzpicturedependsonfile{config_window.tikz}
  \includegraphics[width=\widthTwoThirds]{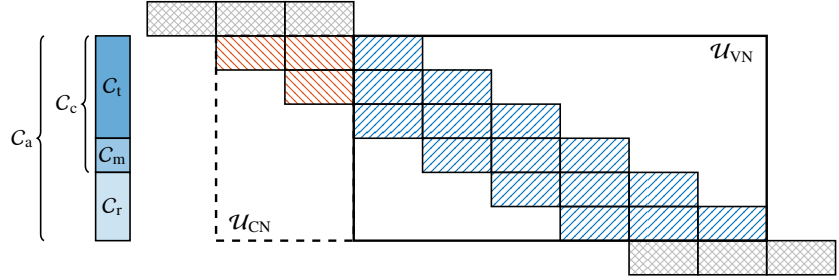}
  \caption{\ac{WD} for $\winSize=6$ and a code with $\mem=2$. The shaded rectangles represent the $\expMat_\idxMem(\idxTime)$. The parts of the exponent matrix for which messages are updated with the \ac{VN}-centered strategy $\cfgVN$ are contained within the thick solid line. With the \ac{CN}-centered strategy $\cfgCN$, the messages for the parts contained within the dashed line are updated as well.}
  \label{fig:config:window}
\end{figure}

\figurename{~\ref{fig:config:window}} depicts the decoding windows for both update strategies assuming \iac{SC}-\ac{LDPC} code with memory $\mem=2$. The shaded rectangles represent the structure of the exponent matrix $\expMat_{[0, \numCouple-1]}$: Each rectangle corresponds to one of the sub-matrices $\expMat_\idxMem(\idxTime)$ with $\numCnSbTime$ \ac{CN} blocks and $\numVnSbTime$ \ac{VN} blocks. Moving the window to the next position means shifting it down and right by one $\expMat_\idxMem(\idxTime)$ each, resulting in a large overlap between windows at two successive positions.

A decoding window with the \ac{VN}-centered strategy $\cfgVN$ is indicated by the thick solid line; the contained sub-matrices are hatched in blue. Performing updates on \acp{VN} in the window still requires access to messages sent along edges related to the previous sub-matrices back-hatched in red, contained within the dashed line. However, those accesses are read-only. By contrast, the messages for both the blue and the red sub-matrices are updated with the \ac{CN}-centered strategy $\cfgCN$. In either case, messages along the edges corresponding to the sub-matrices outside the windows (crosshatched in gray) are not updated.

Please note that the method of amplifying \ac{LLR} magnitudes for read-only messages in case of satisfied \ac{PC} equations in previous window positions as proposed in~\cite{Ali2017} cannot be applied here. That is because we use serial update scheduling throughout this paper, which gives better results but is not compatible with the method of~\cite{Ali2017}.

\subsection{Parity-Based Early Termination and Window-Size Restrictions}

With \ac{PC}-based \ac{ET}, the \ac{WD} moves its decoding window from position $\idxWin$ to position $\idxWin+1$ if a certain number of iterations is completed or all selected \ac{PC} equations are satisfied. We focus on the following sets of \acp{CN} indicated by the vertical bar at the left-hand side of \figurename{~\ref{fig:config:window}}:
\begin{IEEEdescription}[\IEEEsetlabelwidth{$\setCnWinMid$}]
  \item[$\setCnWinTar$] Target \acp{CN}; the top-most $\numCnSbTime(\mem+1)$ blocks of \acp{CN} in a decoding window. Closely related to the so-called target \acp{VN} that are no longer updated when the window moves to the next position.
  \item[$\setCnWinCom$] Complete \acp{CN}; all \acp{CN} in a window except for the bottom-most $\numCnSbTime\mem$ blocks of \acp{CN}. Only connected to \acp{VN} whose edges are completely covered by the decoding window. $\setCnWinCom$ consists of the target \acp{CN} $\setCnWinTar$ and the middle \acp{CN} $\setCnWinMid$.
  \item[$\setCnWinAll$] All \acp{CN} in a decoding window. $\setCnWinAll$ consists of the complete \acp{CN} $\setCnWinCom$ and the remaining \acp{CN} $\setCnWinRem$.
\end{IEEEdescription}
Using the set $\setCnWinAll$ for \ac{ET} is the most obvious approach as it corresponds to the procedure in block-code decoding. The authors of~\cite{Ali2017}, however, propose using the set $\setCnWinTar$.\footnote{The simplest choice, i.e., using only the top-most $\numCnSbTime$ blocks of \ac{PC} equations in the window, would result in a poor decoding performance for $\numCnSbTime=1$ as is used in this paper.} Using the set $\setCnWinCom$, i.e., a choice in-between $\setCnWinTar$ and $\setCnWinAll$, is proposed in~\cite{Kang2018}.

For meaningful comparisons of the various sets of \ac{PC} equations for \ac{ET}, the window size $\winSize$ should be restricted such that $\abs{\setCnWinTar}<\abs{\setCnWinCom}<\abs{\setCnWinAll}$. In other words, we want to exclude the cases $\abs{\setCnWinTar}=\abs{\setCnWinCom}$ and $\abs{\setCnWinCom}=\abs{\setCnWinAll}$. Expressed in terms of code dimensions, these cardinalities are given by
\begin{IEEEeqnarray}{rCl}
  \abs{\setCnWinTar} &=& (\mem + 1) \numCnSbTime\liftFac \\
  \abs{\setCnWinCom} &=& (\winSize - \mem) \numCnSbTime\liftFac \text{~and} \\
  \abs{\setCnWinAll} &=& \winSize\numCnSbTime\liftFac.
\end{IEEEeqnarray}
$\abs{\setCnWinCom}<\abs{\setCnWinAll}$ always holds for $\mem>0$. To achieve $\abs{\setCnWinTar}<\abs{\setCnWinCom}$, we require
\begin{IEEEeqnarray}{rCl}
  \winSize > 2 \mem + 1.
  \label{eq:config:winsize}
\end{IEEEeqnarray}
In general, the window size is kept small to maximize the reduction of internal memory compared to the \ac{FBD}, cf.\ the selected window sizes in~\cite{Iyengar2012,UlHassan2017,Lentmaier2011}, and others. However, a \ac{WD} with a window size smaller than a certain minimal size may have a larger iterative decoding threshold than the \ac{FBD}~\cite{Lentmaier2010}. We still include small window sizes in our evaluations to highlight the effects of nonideal implementations, e.g., caused by fixed-point calculations.

\section{Transmission Setup and Decoder Parameters to Achieve Equal Maximal Complexity}
\label{sec:setup}
\acresetsingle{ET}
\acresetsingle{FBD}

The simulation setup and the parameters used to achieve equal maximal computational complexity are given in this section.

\subsection{Transmission Parameters}

For the simulation results presented below, the (5,10) ensemble as described in Section~\ref{sec:bg} is used. The lifting factor is set to $\liftFac=256$ and the coupling length is $\numCouple=100$, giving a codeword length of $\numVn=51200$. The resulting code rate is $\rate=0.48$, i.e., the rate reduction is 4\% compared to the asymptotic code rate of $\rate_\infty=1/2$. As mentioned above, simulations are performed using only the all-zero codeword. The code symbols are modulated using 16-ary \ac{QAM} with Gray mapping to match the rate proposed for 5G \ac{NR} as given in~\cite{TS38214}; such a setup is also used in~\cite{Schmalen2013}. The code symbols are used in sequence without interleaving before the \ac{QAM} mapping. Scrambling as described in~\cite{TS38211} is used such that the complex-valued \ac{QAM} symbols are uniformly distributed over the whole \ac{QAM} constellation.

A fading channel with \ac{AWGN} is used to simulate transmissions in a typical wireless communication system. The statistical model is taken from~\cite{Muller-Weinfurtner2002} as this model accords well with today's communication systems, e.g., Wi-Fi or 5G \ac{NR}. It accounts for the use of \ac{OFDM} in conjunction with \ac{MIMO} transmissions. The channel is modeled by \ac{MRC} of 4 independent propagation paths, each exhibiting Rayleigh fading. This channel setup corresponds to, e.g., $2\times2$ \acs{MIMO}-\acs{OFDM} with appropriate \ac{SFBC}~\cite{tse:wireless} and ideal \ac{CSI} at the receiver. With sufficient interleaving in frequency, the fading coefficients are independent and identically distributed (i.i.d.) for each \ac{QAM} symbol. The mean \ac{SNR} is indicated by $\esno$ where $\symbolEnergy$ is the average energy per \ac{QAM} symbol and $\psdWhite$ is the one-sided noise \ac{PSD}.

\subsection{General Decoder Configuration}

All decoders in this paper implement the decoding algorithm from~\cite{Jones2003} using fixed-point arithmetic with an \ac{LLR}-magnitude resolution of \SI{10}{\bit}. This decoding algorithm is a blend of the accurate \ac{SPA} and the approximate \ac{MSA}: When processing a \ac{CN}, only the outgoing message sent along the edge with the weakest incoming message is computed accurately with the \ac{SPA}; the combination of \emph{all} incoming messages with appropriate signs (again with the \ac{SPA}) is used for all other outgoing messages.

Furthermore, row-wise serial layered scheduling as described in~\cite{Yeo2001} is applied. With \ac{QC} codes and assuming $\numCnSbTime=1$, the $\liftFac$ \ac{PC} equations formed by one row in $\expMat_{[0, \numCouple-1]}$ are orthogonal to each other, which is why the respective \ac{CN} updates can be processed in parallel without numerical dependencies. One such row is denoted as a \emph{layer}. Thus, the $\idxLayer$-th layer $\layer_\idxLayer$ consists of the following block of $\liftFac$ consecutive \acp{CN}:
\begin{IEEEeqnarray}{rCl}
  \layer_\idxLayer = \left\{ \idxLayer\liftFac, \idxLayer\liftFac+1, \dots, (\idxLayer+1)\liftFac-1 \right\}.
\end{IEEEeqnarray}
\emph{Serial} scheduling means that the outgoing messages are updated for one layer at a time and the related symbol reliabilities are updated before updating the next layer. This procedure stands in contrast to \emph{parallel} update schedules where the messages are updated for all layers before the symbol reliabilities are updated. In our \ac{WD} implementation, the layers within each window are updated from top to bottom with respect to the exponent matrix $\expMat_{[0, \numCouple-1]}$ from~\eqref{eq:bg:pcm_term}.

For consistency, a decoding window of size $\winSize$ is assumed to always contain $\winSize$ layers when $\numCnSbTime=1$. In other words, the top layer in the first window is the top row of the exponent matrix; the bottom layer in the last window is the bottom row of the exponent matrix. Thus, there are
\begin{IEEEeqnarray}{rCl}
  \numWin = \numCouple + \mem - \winSize + 1
  \label{eq:setup:numwin}
\end{IEEEeqnarray}
unique window positions and the window virtually extends beyond the left-hand or right-hand sides of $\expMat_{[0, \numCouple-1]}$ during the first and last $\mem$ positions. These positions contain the \acp{CN} with reduced degrees. Formally, a window of size $\winSize$ at position $\idxWin$, $\idxWin=0\dots(\numWin-1)$, is defined as the set of all layers that are updated:
\begin{IEEEeqnarray}{rCl}
  \win{\winSize}{\idxWin} = \left\{ \layer_\idxLayer : \idxWin \le \idxLayer < \idxWin+\winSize \right\}.
  \label{eq:setup:window}
\end{IEEEeqnarray}

\subsection{Achieving Equal Maximal Complexity}

A \ac{WD} using the \ac{VN}-centered update strategy $\cfgVN$ disregards message updates for the edges back-hatched in red in \figurename{~\ref{fig:config:window}}. The key principle in this paper is to consider this difference in the number of message updates when aiming for equal overall computational complexity for all configurations. To be precise, we measure the complexity in terms of the number of message updates (with respect to the exponent matrix) to decode all $\numInfo$ information symbols. A similar principle is applied in~\cite{Battaglioni2017} to evaluate infinitely long, nonterminated codes. To our best knowledge, comparisons of different decoder configurations with finite-length codes, however, have not yet been proposed.

The upper limit on the decoding complexity is given by \iac{FBD}, i.e., a decoder with $\winSize=\numCouple+\mem$, performing up to $\numIterMax_\text{,FBD}=200$ iterations. We could not find significant improvements in the decoding performance with additional iterations. The \ac{WD} performs an appropriate maximal number of iterations \emph{per window} such that the total maximal number of message updates $\numMUMax$ does not surpass that of the \ac{FBD}. From here on, all quantities regarding the number of message updates relates to the code's exponent matrix, i.e., the lifting factor $\liftFac$ is ignored.

Formally, let $\numMU_{1, \text{FBD}}$ denote the number of message updates with a single iteration with the \ac{FBD} (which equals the number of edges in the Tanner graph):
\begin{IEEEeqnarray}{rCl}
  \numMU_{1, \text{FBD}} = \sum_{\idxCn=0}^{\numCnSb-1} \abs{ \setEdgesVn{\idxCn} }
  \label{eq:setup:numMUOneFBD}
\end{IEEEeqnarray}
where $\numCnSb$ is the number of rows in $\expMat_{[0, \numCouple-1]}$. For a \ac{WD} with the \ac{CN}-centered update strategy $\cfgCN$, the number of message updates with a single iteration in each window is
\begin{IEEEeqnarray}{rCl}
  \numMU_{1,\text{WD}}^{\text{C}}(\winSize) &=& \sum_{\idxWin=0\vphantom{\win{\winSize}{\idxWin}}}^{\numWin-1} \sum_{\idxCn\in\win{\winSize}{\idxWin}} \abs{ \setEdgesVn{\idxCn} }\\
  \intertext{whereas with the \ac{VN}-centered strategy $\cfgVN$ it is}
  \numMU_{1,\text{WD}}^{\text{V}}(\winSize) &=& \sum_{\idxWin=0\vphantom{\win{\winSize}{\idxWin}}}^{\numWin-1} \sum_{\idxCn\in\win{\winSize}{\idxWin}} \abs{ \setEdgesVn{\idxCn} \setminus \setVnWinExt^{(\idxWin)} } \label{eq:setup:numMUOneWD}
\end{IEEEeqnarray}
where $\setVnWinExt^{(\idxWin)}$ denotes the \acp{VN} not updated at position $\idxWin$.

Finally, the maximal computational effort $\numIterMax_\text{,FBD}\numMU_{1, \text{FBD}}$, i.e., the number of message updates when the \ac{FBD} performs $\numIterMax_\text{,FBD}$ iterations, is converted to a maximal number of iterations per window $\numIterMax_\text{,WD}(\winSize)$ for the \ac{WD}:
\begin{IEEEeqnarray}{rCl}
  \numIterMax_\text{,WD}(\winSize) = \floor*{ \numIterMax_\text{,FBD} \frac{\numMU_{1, \text{FBD}}}{\numMU_{1,\text{WD}}(\winSize)} }
\end{IEEEeqnarray}
where $\floor{\cdot}$ is the floor operation.

From here on, we omit subscripts and arguments related to the number of iterations or the number of message updates for brevity. \tablename{~\ref{table:setup:complexity}} lists the maximal number of iterations (per window) $\numIterMax$ and the corresponding maximal number of message updates $\numMUMax=\numIterMax\numMU_{1}$ for all configurations used throughout this paper. The largest relative difference in $\numMUMax$ to the value targeted with the \ac{FBD} is about 5\% due to rounding, which we consider small enough.

\begin{table}[htb]
  \centering
  \begin{tabular}{@{}lrrrr@{}}
    \toprule
    Decoder                            & $\winSize$  & $\numMsgWin$ & $\numIterMax$ &  $\numMUMax$ \\
    \midrule
    \ac{FBD}                           & \textendash &               &           200 & \num{200000} \\
    \midrule
    \multirow{4}{*}{\ac{WD}, $\cfgVN$} &          12 &           100 &            21 & \num{194460} \\
                                       &          14 &           120 &            18 & \num{195840} \\
                                       &          16 &           140 &            16 & \num{198720} \\
                                       &          20 &           180 &            13 & \num{198380} \\
    \midrule
    \multirow{3}{*}{\ac{WD}, $\cfgCN$} &          10 &           100 &            21 & \num{197820} \\
                                       &          12 &           120 &            18 & \num{199440} \\
                                       &          14 &           140 &            15 & \num{189900} \\
    \bottomrule
  \end{tabular}
  \setlength{\abovecaptionskip}{2pt plus 1pt minus 1pt}
  \caption{Number of message updates per window per iteration ($\numMsgWin$), maximal number of iterations per window ($\numIterMax$), and maximal number of message updates with $\numIterMax$ iterations per window ($\numMUMax$).}
  \label{table:setup:complexity}
\end{table}

Instead of comparing $\cfgVN$ and $\cfgCN$ for the same window size and allowing different values for $\numIterMax$, we choose different window sizes such that the number of message updates per window per iteration (described by $\numMsgWin$ in \tablename{~\ref{table:setup:complexity}) is roughly the same for both configurations.\footnote{The values for $\numMsgWin$ assume a single decoding window in the middle of the exponent matrix. By contrast, the remaining values in \tablename{~\ref{table:setup:complexity}} assume a terminated code with $\numCouple=100$ as described above.} Even though the match is not always perfect, a \ac{WD} using $\cfgCN$ with a window of size $\winSize$ is compared to a \ac{WD} using $\cfgVN$ with a window of size $\winSize+2$ in our simulations presented in Section~\ref{sec:results}.

\subsection{Metrics for Evaluation}

\acused{BLER}
\acused{ANMU}
The decoding performance is evaluated by means of the \acfi{BLER} (cf.~\eqref{eq:bg:info_parity}\textendash \eqref{eq:bg:info_parity_est})
\begin{IEEEeqnarray}{rCl}
  \bler \coloneqq \Pr\{ \vec{\hat{\infoBit}} \ne \vec{\infoBit} \}
  \label{eq:setup:bler}
\end{IEEEeqnarray}
and the \acfi{ANMU} $\aneu$ with respect to the exponent matrix. With \ac{ET}, $\aneu\le\numMUMax$. Due to the differences in $\numMUMax$ across different configurations, we evaluate the ratio
\begin{IEEEeqnarray}{rCl}
  \aneuRel \coloneqq \frac{\aneu}{\numMUMax}
  \label{eq:setup:aneu_rel}
\end{IEEEeqnarray}
and refer to $\aneuRel$ as the \emph{relative} \ac{ANMU}.

\section{Joint Evaluation of Early Termination and Update Strategies}
\label{sec:results}
\acresetsingle{ANMU}
\acresetsingle{BLER}
\acresetsingle{ET}
\acresetsingle{FBD}

The numerical results in this section have been obtained from $\num{1e6}$ realizations of the channel and the code ensemble each, i.e., a new code realization and a new channel realization have been generated for each new transmission. The simulation setup is described in Section~\ref{sec:setup}. Our aim is to determine the configuration for a \ac{WD} with the smallest resulting average computational complexity without sacrificing the decoding performance too much. We jointly consider \ac{ET} based on \ac{PC} equations, different window update strategies, and various window sizes $\winSize$ as discussed in Section~\ref{sec:config}. To keep the evaluation organized, we first present results for varying window sizes $\winSize$ and various sets of \ac{PC} equations for \ac{ET} using only the \ac{VN}-centered update strategy $\cfgVN$. Afterwards, we present a comparison of $\cfgVN$ against the \ac{CN}-centered update strategy $\cfgCN$. In all cases, the maximal number of iterations per window was adjusted for equal maximal complexity according to \tablename{~\ref{table:setup:complexity}}. The results for \iac{FBD} are given for reference.

\subsection{Impact of the Window Size and Early Termination}

For the first part of this evaluation, we selected the \ac{CN} sets $\setCnWinCom$ (complete \acp{CN}) and $\setCnWinTar$ (target \acp{CN}) for \ac{ET}. The \ac{CN} set $\setCnWinAll$ (all \acp{CN}) was excluded since \ac{PC} equations at the bottom of a window are expected to almost never be satisfied. That is because the \acp{BER} of the right-most \acp{VN} in the window are close to 1~\cite{UlHassan2017}. We selected window sizes $\winSize\in\curly{12,16,20}$ to satisfy the constraint given by~\eqref{eq:config:winsize} and to get an overview of larger windows as well.

\figurename{~\ref{fig:results:winet_aneu}} portrays the relative \ac{ANMU} $\aneuRel$ from~\eqref{eq:setup:aneu_rel} over the \ac{SNR} $\esno$. Only considering the \ac{PC} equations in $\setCnWinTar$ for \ac{ET} reduces $\aneuRel$ compared to considering the \ac{PC} equations in $\setCnWinCom$. The top-most \ac{PC} equations in a window are formed by more reliable \acp{VN}. In addition, the smaller number of \ac{PC} equations in $\setCnWinTar$ can become satisfied more quickly. Depending on the \ac{SNR}, $\aneuRel$ is reduced by up to 49\% with $\winSize=12$; up to 58\% with $\winSize=16$, and by up to 60\% with $\winSize=20$ when selecting the set $\setCnWinTar$ instead of $\setCnWinCom$ for \ac{ET}. Moreover, any \ac{WD} evaluated here converges faster than the \ac{FBD} except when considering the domain of very high \ac{SNR}.

\begin{figure}[htb]
  \centering
  \tikzsetnextfilename{winet_aneu}
  \tikzpicturedependsonfile{winet_aneu.tikz}
  \includegraphics[width=\widthTwoThirds,axisratio=\plotratio]{winet_aneu}
  \caption{Relative \acs*{ANMU} $\aneuRel$ over the \ac{SNR} $\esno$ with various sets of \ac{PC} equations for \ac{ET} and various window sizes $\winSize$ using the update strategy $\cfgVN$.}
  \label{fig:results:winet_aneu}
\end{figure}

The window size $\winSize$ also influences $\aneuRel$: When selecting the \ac{PC} equations in $\setCnWinCom$ for \ac{ET}, choosing smaller windows results in faster convergence than choosing larger windows. Since $\abs{\setCnWinCom}$ grows with the window size, larger windows require more \ac{PC} equations to be satisfied. This, in turn, requires a greater amount of message updates. When choosing $\setCnWinTar$ for \ac{ET}, however, the differences between smaller and larger window sizes are much smaller because $\abs{\setCnWinTar}$ is independent of $\winSize$. In this case, a large window may actually \emph{increase} the decoding complexity because more messages are updated in each iteration. This effect is especially visible in the domain of high \ac{SNR} where usually a single iteration per window is sufficient to satisfy the \ac{PC} equations in $\setCnWinTar$.

The decoding performance as measured by the \ac{BLER} $\bler$ from~\eqref{eq:setup:bler} is mostly determined by the window size $\winSize$. As shown in \figurename{~\ref{fig:results:winet_cber}}, using larger windows results in fewer residual errors because of the simplified transport of information across the whole Tanner graph. At a working point of $\bler=\num{1e-2}$, the gap in the \ac{SNR} to the \ac{FBD} constitutes around \SI{0.25}{\decibel} with any $\winSize$. The gap increases with increasing \ac{SNR} where the largest gap can be observed for $\winSize=12$; the differences between the decoders with $\winSize=16$ and $\winSize=20$ are negligible. Interestingly, the choice of \ac{PC} equations for \ac{ET} only has a minor impact on $\bler$: Using the smaller set $\setCnWinTar$ results in about the same (or even a smaller) number of residual decoding errors as using the larger set $\setCnWinCom$. Thus, the smaller number of \ac{PC} equations in $\setCnWinTar$ seems sufficient to verify the estimates of the sub-codewords in the decoding window.

\begin{figure}[htb]
  \centering
  \tikzsetnextfilename{winet_cber}
  \tikzpicturedependsonfile{winet_cber.tikz}
  \includegraphics[width=\widthTwoThirds,axisratio=\plotratio]{winet_cber}
  \caption{\acs*{BLER} $\bler$ over the \ac{SNR} $\esno$ with various sets of \ac{PC} equations and various window sizes $\winSize$ using the update strategy $\cfgVN$.}
  \label{fig:results:winet_cber}
\end{figure}

Furthermore, the error floor is relatively high\textemdash even for the \ac{FBD}. Despite the omission of code realizations with cycles of length 4 in the Tanner graph, the employed code ensemble does not perform well in general. The \ac{WD}'s error floor is even higher than the \ac{FBD}'s since the \ac{WD} can easily get stuck and propagate errors due to the unidirectional decoding. Nonetheless, our main concern is the gap in the \ac{SNR} at $\bler=\num{1e-2}$, which is a typical working point in cellular systems~\cite{TS381011}.

Summing up, \ac{ET} based on \ac{PC} equations works very well. The set $\setCnWinTar$ represents the best choice for \ac{ET} in all considered aspects as a very low computational complexity can be achieved this way. Additionally, a \ac{WD} with a surprisingly large window size of, e.g., $\winSize=16$ seems to be an overall good choice as it pratically matches the decoding performance of a \ac{WD} with $\winSize=20$ while offering a reduced computational complexity.

\subsection{Comparison of Update Strategies}

For the second part of this evaluation, we compared the two window update strategies reviewed in Section~\ref{sec:config}. Specifically, we compared a \ac{WD} using $\cfgCN$ with window size $\winSize$ against a \ac{WD} using $\cfgVN$ with window size $\winSize+2$ as argued in Section~\ref{sec:setup}. Furthermore, we selected $\winSize\in\curly{12,14,16}$ for $\cfgVN$ and thus $\winSize\in\curly{10,12,14}$ for $\cfgCN$ as larger windows did not significantly improve the decoding performance, cf.\ \figurename{~\ref{fig:results:winet_cber}}. Following our findings from the evaluations above, we solely considered the \ac{CN} set $\setCnWinTar$ for \ac{ET}.

\figurename{~\ref{fig:results:winup_5_10_aneu}} depicts the relative \ac{ANMU} $\aneuRel$ for the configurations listed above. Using $\cfgCN$ with $\winSize=10$ results in the worst decoding complexity. All other configurations perform slightly better with only small differences among them. Using $\cfgVN$ results in minor advantages over using $\cfgCN$. As before, all \ac{WD} configurations beat the \ac{FBD} in terms of computational complexity.

\begin{figure}[htb]
  \centering
  \tikzsetnextfilename{winup_5_10_aneu}
  \tikzpicturedependsonfile{winup_5_10_aneu.tikz}
  \includegraphics[width=\widthTwoThirds,axisratio=\plotratio]{winup_5_10_aneu}
  \caption{Relative \ac{ANMU} over the \ac{SNR} $\esno$ comparing $\cfgCN$ with $\winSize\in\curly{10,12,14}$ against $\cfgVN$ with $\winSize\in\curly{12,14,16}$. The \ac{PC} equations in $\setCnWinTar$ are used for \ac{ET}.}
  \label{fig:results:winup_5_10_aneu}
\end{figure}

The \ac{BLER} $\bler$ is shown in \figurename{~\ref{fig:results:winup_5_10_cber}}. According with the results from \figurename{~\ref{fig:results:winet_cber}}, the window size $\winSize$ is the most decisive factor for the resulting decoding performance. Still, the differences between using $\cfgVN$ with window size $\winSize+2$ and using $\cfgCN$ with window size $\winSize$ are negligible. Thus, the comparison with a normalized per-window complexity can indeed be considered to be fair.

\begin{figure}[htb]
  \centering
  \tikzsetnextfilename{winup_5_10_cber}
  \tikzpicturedependsonfile{winup_5_10_cber.tikz}
  \includegraphics[width=\widthTwoThirds,axisratio=\plotratio]{winup_5_10_cber}
  \caption{\ac{BLER} over the \ac{SNR} $\esno$ comparing $\cfgCN$ with $\winSize\in\curly{10,12,14}$ against $\cfgVN$ with $\winSize\in\curly{12,14,16}$. The \ac{PC} equations in $\setCnWinTar$ are used for \ac{ET}.}
  \label{fig:results:winup_5_10_cber}
\end{figure}

Summing up the whole evaluation, using only the \ac{PC} equations corresponding to the \acp{CN} in $\setCnWinTar$ is the superior choice for minimizing the computational complexity. The computational complexity is then reduced when using a medium window size (e.g., $\winSize=16$) rather than smaller or larger window sizes. Furthermore, the \ac{VN}-centered update strategy $\cfgVN$ is the overall better update strategy: The computational complexity is improved while the decoding performance is kept at a similar level as a window of size $\winSize+2$ can be used in a fair comparison against $\cfgCN$ with window size $\winSize$. The only downside of using $\cfgVN$ is the more challenging implementation in conjunction with \ac{CN}-wise serial scheduling and the \acf{MSA} since not all messages for all the \acp{CN} in the decoding window are updated.

\section{Conclusion}
\label{sec:conclusion}
\acresetsingle{FBD}
\acresetsingle{WD}

Limiting a \ac{WD} for \ac{SC}-\ac{LDPC} codes to the same maximal decoding complexity as \iac{FBD} reveals the following results: It is only possible to beat the \ac{FBD} in terms of computational complexity with an optimized configuration. With the configuration derived in this paper, the resulting \ac{WD} requires on average only half the decoding complexity of the \ac{FBD} while simultaneously showing a gap in the \ac{SNR} of only around \SI{0.25}{\decibel} for a working point with \ac{BLER} $\bler=\num{1e-2}$. Comparison of various update strategies shows that it may be advantageous to skip updates of some messages in the decoding window. The window can then be enlarged to achieve the same computational complexity while improving the decoding performance. Lastly, all results obtained in this paper have been achieved with configurations requiring low control overhead, making the proposed \ac{WD} design very suitable for efficient hardware implementation.

\bibliographystyle{elsarticle-num}
\bibliography{IEEEabrv,ms}

\end{document}